\documentclass[aps,prl,twocolumn]{revtex4}
\usepackage{amsmath,bm,epsfig}
\def\Fbox#1{\vskip1ex\hbox to 8.5cm{\hfil\fboxsep0.3cm\fbox{%
  \parbox{8.0cm}{#1}}\hfil}\vskip1ex\noindent}  %%  {TEXT} in BOX

\newcommand{\B}[1]{{\bm{#1}}}%% Bold Roman & Greek Lower & Upper Case
\let \= \equiv \let\*\cdot \let\~\widetilde \let\^\widehat \let\-\overline

\def\<{\left\langle}    \def\>{\right\rangle}
\def\({\left(}          \def\){\right)}
 \def \[ {\left [} \def \] {\right ]}

\begin{document}
\title{Relations between the Material Mechanical Parameters and the Inter-particle Potential in Amorphous Solids}
\author{Edan Lerner$^{1}$ and Itamar Procaccia$^{1,3}$}
\affiliation{$^1$Department of Chemical Physics,
The Weizmann Institute of Science, Rehovot 76100, Israel}
\author{Emily S.C. Ching$^{2,3}$}
\affiliation{$^2$Department of Physics, The Chinese University of Hong Kong, Shatin, Hong Kong}
\affiliation{$^3$Institute of Theoretical Physics, The Chinese University of Hong Kong, Shatin, Hong
Kong }
\author{H.G.E Hentschel}
\affiliation{Dept. of Physics, Emory University, Atlanta Ga. 30322}
\date{\today}

 \begin{abstract}
 The shear-modulus and yield-stress of amorphous solids are important material parameters,
with the former determining the rate of increase of stress under external strain and the latter being the
stress value at which the material flows in a plastic manner. It is therefore important to understand
how these parameters can be related to the inter-particle potential. Here a scaling theory is
presented such that given the inter-particle potential, the dependence of the yield stress and the shear modulus
on the density of the solid can be predicted in the athermal limit. It is explained when such prediction is possible
at all densities and when it is only applicable at high densities.
These results open up exciting possibilities for designing in principle new materials with desirable mechanical properties.
 \end{abstract}
  \maketitle

{\bf Introduction}:
Typical solids, whether amorphous or crystalline, respond linearly and elastically to small strains.
Then the stress $\sigma_{ij}$ increases like $\mu\epsilon_{ij}$ where $\mu$ is the shear modulus, $\epsilon_{ij}$ is the strain
 \begin{equation}
 \epsilon_{ij} \equiv \frac{1}{2}\left(\frac{\partial u_i}{\partial x_j}+\frac{\partial u_j}{\partial x_i}\right) \ ,
 \end{equation}
 and $\B u$ is the displacement field. Upon increasing the strain further the material yields and begins to flow; the average stress-value
 in the plastic steady state is defined here as the ``yield-stress" $\sigma_Y$. In crystalline materials this plastic flow is understood in
terms of the motion of defects like dislocations \cite{66ST}. It is much less clear in amorphous solids what the carriers of plasticity are
and how to relate the onset of plasticity to the microscopic properties of the material.
The aim of this Letter is to present a scaling theory
  that is able, subject to stated conditions, to predict how $\mu$ and $\sigma_Y$ depend, say, on the density of the material. In addition
   we clarify the issue of universality of the ratio $\mu/\sigma_Y$ which had come up in recent experiments on a family of metallic glasses
   \cite{05JS}.

   %%%%%%%%%%%%%%%%%%%%%
\begin{figure}
\centering
\includegraphics[scale = 0.5]{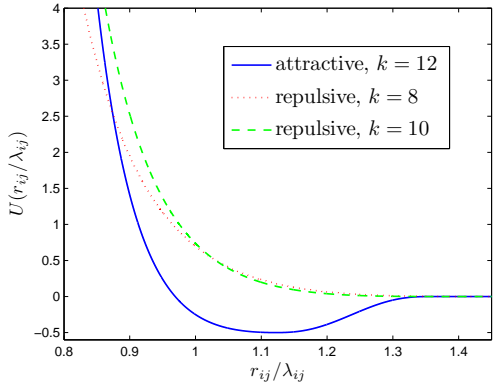}
\caption{Color online: The three different potential discussed in this Letter.}
\label{3pots}
\end{figure}
%%%%%%%%%%%%%%%%%%%%%%%%%%%%%%%%%%%

\begin{figure}
\centering
\includegraphics[scale = 0.4]{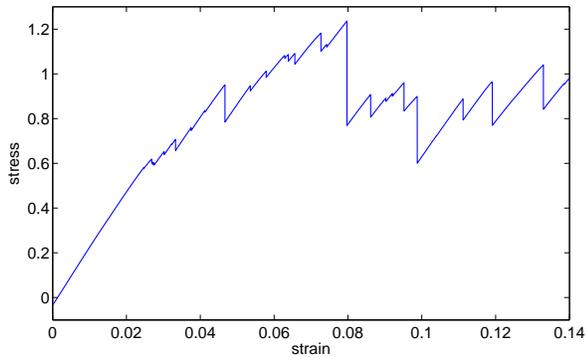}
\caption{ Typical stress-strain curves for a system with $N=4096$ and $k=10$ in Eq. (\ref{potential}).}
\label{strstr}
\end{figure}

 {\bf The yields stress and the energy drops}. To initiate the discussion we remind the readers
of a typical strain-stress curve in amorphous materials. For this purpose we construct a glassy system consisting of poly-dispersed soft disks.
We work with $N$ point particles of equal mass $m$ in two dimensions with pair-wise interaction potentials. Each particle $i$ is assigned an
interaction parameter $\lambda_i$ from a normal distribution with mean $\langle \lambda \rangle = 1$.
The variance is governed by the poly-dispersity parameter $\Delta = 15\%$ where $\Delta^2 = \frac{\langle \left(\lambda_i -
\langle \lambda \rangle \right)^2\rangle}{\langle \lambda \rangle^2} $.
With the definition $\lambda_{ij} = \case{1}{2}(\lambda_i + \lambda_j)$
the potential is constructed such as to minimize computation time \cite{09LP}
\begin{widetext}
\begin{equation}
U(r_{ij}) =
\left\{
\begin{array}{ccl}
\!\!\! \epsilon\left[\left(\frac{\lambda_{ij}}{r_{ij}}\right)^{k}\!\! -\!\!\frac{k(k+2)}{8}
\left( \frac{B_0}{k} \right)^{\frac{k+4}{k+2}}\left(\frac{r_{ij}}{\lambda_{ij}}\right)^4
+ \frac{B_0(k+4)}{4}\left(\frac{r_{ij}}{\lambda_{ij}}\right)^2
-\frac{(k+2)(k+4)}{8}\left( \frac{B_0}{k} \right)^{\frac{k}{k+2}}\right] & , & r_{ij} \le
\sigma_{ij}\left( \frac{k}{B_0} \right)^{\frac{1}{k+2}} \\
0 & , & r_{ij} >
\sigma_{ij}\left( \frac{k}{B_0} \right)^{\frac{1}{k+2}}
\label{potential}
\end{array}
\right\}\ ,
\end{equation}
\end{widetext}
where $r_{ij}$ is the separation between particles $i$ and $j$.
The shape of this potential for $k=8$ and $k=10$ is shown in Fig. \ref{3pots}. Below the units of length, energy, mass and temperature are $\langle \sigma\rangle$, $\epsilon$, $m$ and $\epsilon/k_B$ where
$k_B$ is Boltzmann's constant. The unit of time $\tau_0$ is accordingly $\tau_0=\sqrt{(m\langle \sigma\rangle^2/\epsilon})$.
 In the present simulations we choose $B_0=0.2$ and employ an athermal quasi-static scheme \cite{09LP} with Lees-Edwards boundary conditions.  In Fig. \ref{strstr} we show a typical stress-strain
%%%%%%%%%%%%%%%%%%%%%
%%%%%%%%%%%%%%%%%%%%%%%%%%%%%%%%%%%
curve for this model with $N=4096$ and $k=10$. For small strains the stress grows linearly with a slope $\mu$;
this continues as long as the material is elastic, i.e. for stress less than $\sigma_Y$. For
stress exceeds $\sigma_Y$, plastic events begin to take place. In this regime
we observe drops in the stress of size $\Delta_\sigma$, and we denote the mean size of these stress drops by $\langle \Delta_\sigma \rangle$.
Between the stress drops the strain
 is increased by intervals $\Delta_\epsilon$ with a mean $\langle \Delta_\epsilon\rangle$.
Associated with the stress drops
 are potential energy drops which we denote by $\Delta_U$ whose mean is $\langle \Delta_U\rangle$.
 In \cite{09LP} the dependence of these averages on the system size was determined, with the results
\begin{equation}
\langle \Delta_\sigma \rangle = C_\sigma N^{\beta}\ , \quad \langle \Delta_U\rangle=C_U N^\alpha\ ,
\label{coeff}
\end{equation}
with $C_\sigma$ having the units of stress and $C_U$ of energy. The exponents satisfy a scaling relation
$\alpha-\beta=1$ which can be easily derived from the exact energy conservation relation \cite{06ML,09LP}
\begin{equation}
\sigma_Y \times\langle \Delta_\epsilon \rangle \times V =
\sigma_Y\times \frac{\langle\Delta_\sigma\rangle }{\mu}\times  V =\langle \Delta_U  \rangle \ .
\label{energy}
\end{equation}
Substituting Eq.~(\ref{coeff}) into Eq.~(\ref{energy}),
the experimental fact that $\sigma_Y$ and $\mu$ are intensive quantities independent of $N$ implies $\alpha-\beta=1$,
and we can rewrite the last equation in favor of the yield stress $\sigma_Y$,
\begin{equation}
\sigma_Y = \frac{\rho ~ C_U \mu}{C_\sigma} \ .
\end{equation}
where $\rho\equiv N/V$ is the number density,
Note that the combination $\Omega^{-1}\equiv \rho C_U/C_\sigma$ is a pure number whose non-universality will be discussed below.
The challenge at this point is to provide a theory for the dependence of $\mu$ and the pre-factors $C_\sigma$ and $C_U$ on the experimental control parameters. In this Letter we focus on the athermal limit $T=0$ and study the density dependence; the thermal theory is
more involved and will be dealt with in a future publication.
%%%%%%%%%%%%%%%%%%%%%
\begin{figure}
\centering
\includegraphics[scale = 0.5]{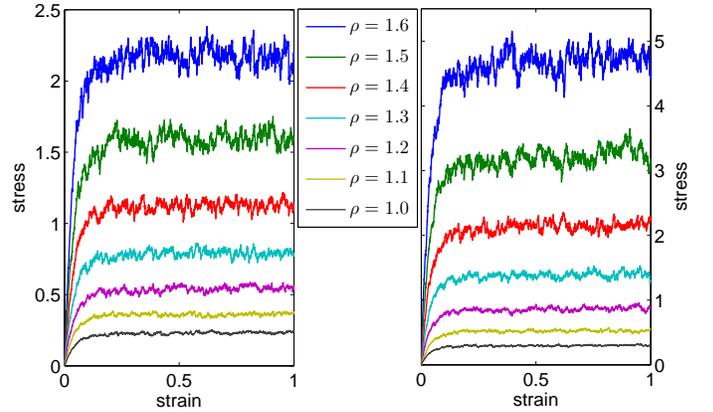}
\caption{Color online: stress-strain curves averaged over 20 independent runs for an athermal system with $N=4096$, $k=8$ (left panel) and $k=10$ (right panel) as a function of the density, with the density increasing from bottom to top.}
\label{rawsig}
\end{figure}
%%%%%%%%%%%%%%%%%%%%%%%%%%%%%%%%%%%%%

\begin{figure}
\centering
\includegraphics[scale = 0.43]{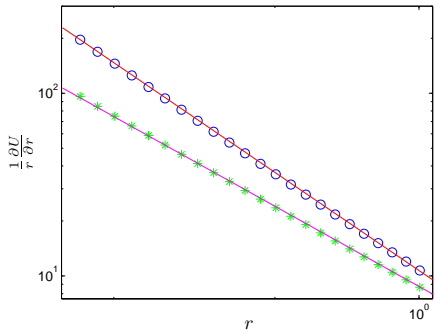}
\caption{Color online: $r^{-1} \frac{\partial U(r)}{\partial r}$ in the range of $r\in [\rho_{\rm max}^{-1/2},\rho_{\rm min}^{-1/2}]$. The line through the points represents the scaling laws (\ref{sigmu})}
\label{scaling}
\end{figure}

\begin{figure}
\centering
\includegraphics[scale = 0.48]{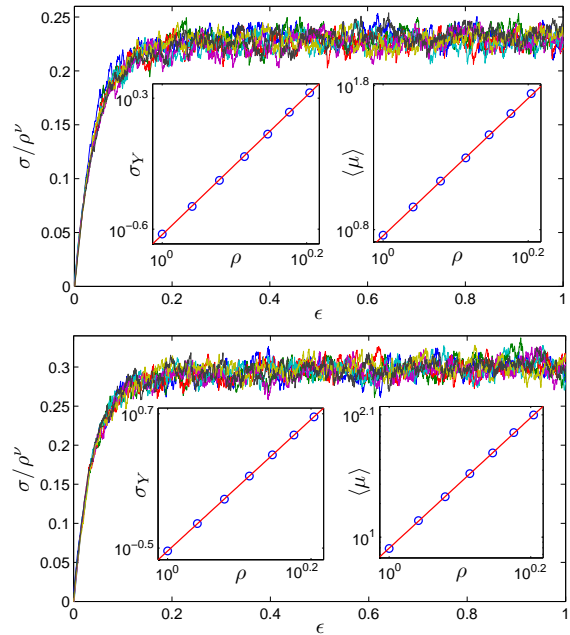}
\caption{Color online: The same stress-strain curves as in Fig. \ref{rawsig} but with the stress rescaled by $\rho^\nu$, with
$\nu=4.80$ for $k=8$ and $\nu=5.87$ for $k=10$. The insets demonstrate the density dependence of $\sigma_Y$ and $\mu$ according
to $\rho^\nu$. }
\label{scalesig}
\end{figure}

 {\bf Results and Scaling Theory}: In Fig. \ref{rawsig} we present the stress-strain curves of two systems, each with $N=4096$,
as a function of the density, one with $k=8$ in Eq. (\ref{potential}) and one
with $k=10$. We note the very strong dependence of both $\mu$ and $\sigma_Y$ on the density (60\% change in the density results in a factor of up to 20 in $\sigma_Y$). We also note the sensitivity to the
 repulsive law of the potential, changing $k$ by 25\% changes the yield stress by a factor of 2. To understand this density dependence
 we point out that both $\sigma_Y$ and $\mu$ have the dimensions of stress, and from the potential the quantity having the same dimension is
%therefore we compute for our potentials
the function $r^{-1} \frac{\partial U(r)}{\partial r}$. Thus if this function has a scaling behavior with $r$ then $\sigma_Y$ and $\mu$
would have a related scaling behavior with $\rho$.
If $U(r)$ were a simple power law $r^{-k}$ this function would behave
like $r^{-k-2}$ or scale with $\rho$ as $\rho^{(k+2)/2}$ since $\rho\sim r^{-2}$.
The potential Eq. (\ref{potential}) is not a simple power law
but as we shall see the function $r^{-1} \frac{\partial U(r)}{\partial r}$ exhibits an effective power law or scaling behavior.
We compute $r^{-1} \frac{\partial U(r)}{\partial r}$ in the range of
$r\in [\rho_{\rm max}^{-1/2},\rho_{\rm min}^{-1/2}]$, see Fig. \ref{scaling}.
 %%%%%%%%%%%%%%%%%%%%%
%%%%%%%%%%%%%%%%%%%%%%%%%%%%%%%%%%%%%
We find that to a very good approximation
 \begin{equation}
\case{1}{r}\frac{\partial U(r)}{\partial r} \sim r^{-2\nu} \ ,
\label{fundsc}
 \end{equation}
with $\nu=4.80$ for $k=8$ and $\nu=5.87$ for $k=10$.
With this effective scaling behavior,
we can predict {\em a-priori} that $\Omega^{-1}$, being a number, should be independent of $\rho$ and
both $\sigma_Y$ and $\langle \mu\rangle$ should scale like
 \begin{equation}
 \sigma_Y\sim \rho^\nu\ , \quad \langle\mu\rangle\sim\rho^\nu \ , \label{sigmu}
 \end{equation}
  where $\langle \mu \rangle$ denotes an average of the shear modulus over the elastic steps of the steady state plastic flow.
  Consequently we expect that the stress-strain curves would collapse by dividing all the stresses by $\rho^\nu$. Fig. \ref{scalesig}
demonstrates the perfect data collapse under this rescaling, together with the high precision of the scaling laws~(\ref{sigmu}).
%%%%%%%%%%%%%%%%%%%%%
%%%%%%%%%%%%%%%%%%%%%%%%%%%%%%%%%%%%%

\begin{figure}
\centering
\includegraphics[scale = 0.45]{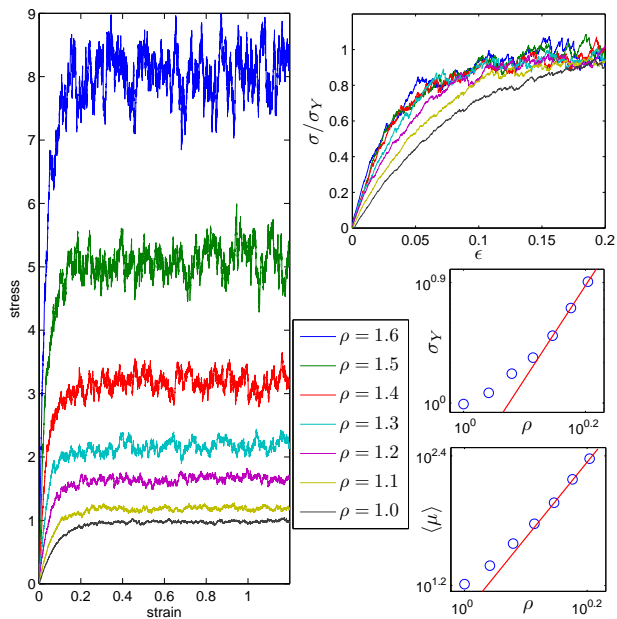}
\caption{Color online. Left panel: stress-strain curves for the potential (\ref{newpot}) which has a repulsive and an attractive part. Right upper panel: demonstration of the failure of rescaling of the stress-strain curves. Lower panels: $\sigma_Y$ and $\langle \mu\rangle $ as a function of the density.  Note that predictability is regained only for higher densities, the straight line is $\rho^7$.}
\label{ststnew}
\end{figure}

{\bf When scaling fails}: To delineate the applicability of the scaling theory,
 we turn now to a third potential in which to the repulsive branch we add an attractive one, see Fig. \ref{3pots}.
The potential chosen reads (with the same polydispersity of $\lambda_{ij}$)
\begin{equation}
U(r) = \left\{
\begin{array}{ccl}
\tilde{U}(r_{ij})  & , & r \le r_0(\lambda_{ij}) \\
\hat{U}(r_{ij})  & , & r_0(\lambda_{ij}) < r  \le r_c(\lambda_{ij}) \\
0 & , & r > r_c(\lambda_{ij})
\end{array}
\right.
\label{newpot}
\end{equation}
with $\tilde{U}(r_{ij}) = \epsilon\left[\left(\frac{\lambda_{ij}}{r_{ij}}\right)^k -
\left(\frac{\lambda_{ij}}{r_{ij}}\right)^6 - 1/4 \right]$; $k=12$, $r_0 = 2^{1/6} \lambda_{ij}$ and $r_c=1.36\lambda_{ij}$. The attractive part $\hat U(r)$ is
glued smoothly to the repulsive part. We choose  $\hat U(r) =\frac{\epsilon}{2} P\left(\frac{r-r_0}{r_c-r_0}\right)$ where $P(x) = \sum_{i=0}^5A_ix^i$  and the coefficients $A_i$ are chosen such that the potential is smooth up to second derivative \cite{params}.
We repeated the measurements of the stress-strain curves for this system, now with $N=2500$, and display the results in
Fig. \ref{ststnew}, left panel. In the right upper panel we show what happens when we try to collapse the data by rescaling
the stress by $\sigma_Y$. Of course the stress-stain curves now all asymptote to the same value, but the curves fail to collapse,
since $\mu$ does not in general scale in the same way as $\sigma_Y$. The reason is the failing of the scaling hypothesis in this
case: the calculation of $r^{-1} \frac{\partial U(r)}{\partial r}$ in the range of $r\in [\rho_{\rm max}^{-1/2},\rho_{\rm min}^{-1/2}]$
fails to provide an effective power law, destroying the scaling hypothesis and the data collapse. Nevertheless
%%%%%%%%%%%%%%%%%%%%%
%%%%%%%%%%%%%%%%%%%%%%%%%%%%%%%%%%%%%
even in the present case we can have predictive power for high densities. When the density increases the repulsive part of the
potential (\ref{newpot}) becomes increasingly more relevant, and the inner power law $r^{-12}$ becomes dominant. We therefore
expect that for higher densities scaling will be regained, and both $\sigma_Y$ and $\langle \mu \rangle$ would depend on the density as $\rho^7$.
The two lower right panels in Fig. \ref{ststnew} show how well this prediction is realized also in the present case.

We therefore conclude that given a potential one should be able to decide whether a scaling theory should be applicable and at what
densities. Whenever the most probable inter-particle distance agrees with the average inter-particle distance scaling can be
employed with impunity. At low densities this condition may not apply due to the attractive part of the potential. Nevertheless with higher densities
scaling and predictability of the mechanical properties should get better and better.

\begin{figure}
\centering
\includegraphics[scale = 0.45]{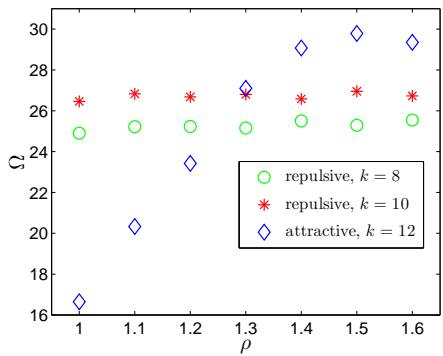}
\caption{Color online: The pure number $\Omega$ as a function of the density for the three potentials discussed in the text. Note that $\Omega$
appears to increase with the exponent of the repulsive part of the potential whenever scaling prevails.}
\label{Omega}
\end{figure}

Finally we discuss the numerical value of the parameter $\Omega=C_\sigma/(\rho C_U)$.  We note that the data collapse
indicates that this parameter is, to a very good approximation, independent of $\rho$ for a given potential function.
For the two different potentials (\ref{potential}) with $k=8$ and $k=10$ respectively, we find from our numerics that
this parameter differs by about 5\%, indicating non-universality. The lack of universality is even clearer
with the last potential (\ref{newpot}). In Fig. \ref{Omega} we
%%%%%%%%%%%%%%%%%%%%%
%%%%%%%%%%%%%%%%%%%%%%%%%%%%%%%%%%%%%%%%%%%%%%%%%%%%%%%%%%
display the computed values of $\Omega$ for the three cases studied above. When scaling prevails the value of $\Omega$
is constant up to numerical fluctuations. In the third case, where scaling fails, $\Omega$ is a strong function of $\rho$ except
at higher densities where scaling behavior is recaptured as explained. We can therefore conclude that the approximate constancy
of $\Omega$ found in a family of metallic glasses \cite{05JS}, is not fundamental but only an indication of the similarity
of the potentials for this family. In general $\Omega$ can depend on the inter-particle potential.
It is quite clear from considering Eqs. (\ref{fundsc}) and ({\ref{sigmu})
that the {\em coefficients} in the scaling laws (\ref{sigmu}) may well depend on the exponent $k$ in the
repulsive part of the potential. The ratio of these pre-factors, being a pure number, could be
independent of $k$, and $\Omega$ could be universal. It appears however that $\mu$ is increasing more
with $k$ than $\sigma_Y$, and $\Omega$ shows a clear increase upon increasing $k$.
How to explain this dependence is beyond our present state of understanding and must remain
an interesting riddle for future research.

In summary, we have begun in this Letter to explore the relations between inter-particle potentials and the mechanical
properties of amorphous solids that are made of these particles. Presently we focused on the athermal limit and showed how
the dependence of the yields stress and the shear modulus on the density can be determined from first principles as long
as the scaling hypothesis prevails. We also examined when this hypothesis failed, and pointed out that in any material
the theoretically determined relations can be trusted at high densities. In later publications we will examine the
effect of temperature on the present issues.

\acknowledgments

This work had been supported in part by the Israel Science Foundation, the German Israeli Foundation, and the Minerva Foundation,
Munich, Germany.

\end{document}